\documentclass[twocolumn,pre,showpacs,amsmath,longbibliography]{revtex4-2}
\usepackage{graphicx}
\usepackage{ascmac}
\usepackage{amsmath}
\usepackage{amssymb}
\usepackage[version=3]{mhchem}
\usepackage{physics}

\begin{document}

\title{Discreteness-induced spatial chaos versus fluctuation-induced spatial order in stochastic Turing pattern formation}
\author{Yusuke Yanagisawa}
\affiliation{Department of Physics, Kyoto University, Kyoto 606-8502, Japan}
\author{Shin-ichi Sasa}
\affiliation{Department of Physics, Kyoto University, Kyoto 606-8502, Japan}
\date{\today}

\begin{abstract}
    We investigate Turing pattern formation in a stochastic reaction-diffusion model defined on $N$ lattice sites,
    where each lattice site is associated with a reaction vessel of volume $\Omega$.
    We focus on a regime where spatial discreteness plays a crucial role, namely when the characteristic length of patterns is comparable to the lattice spacing. 
    In this setting, we compare two different limiting procedures and show that they lead to qualitatively different outcomes. 
    If we first take the deterministic limit $\Omega \to \infty$ and then the long-time limit $t \to \infty$, 
    the stationary solutions of the corresponding spatially discrete deterministic equations become spatially chaotic in the limit $N\to\infty$. 
    In contrast, if we first take the limit $t \to \infty$ and then take an appropriate limit of $\Omega \to \infty$ and $N\to\infty$, the resulting patterns are spatially periodic.
\end{abstract}

\maketitle

\section{Introduction}

% Intro #1
Pattern formation phenomena are widely observed in a variety of macroscopic systems, 
including reaction-diffusion systems~\cite{nicolis1977self,murray2002mathematical}, 
fluid dynamics~\cite{langer1980instabilities,bodenschatz2000recent}, 
soft matter~\cite{bates1990block, seul1995domain, buka1996pattern}, 
active matter~\cite{romanczuk2012active,marchetti2013hydrodynamics,cates2015motility,saha2020scalar}, 
and biological systems~\cite{koch1994biological,rietkerk2008regular,kondo2010reaction}. 
To explain various characteristic patterns, a number of theories that do not rely on the details of the constituents have been proposed.
One of the most successful theoretical frameworks is Turing's theory for reaction-diffusion systems~\cite{turing1952the}, 
where diffusion destabilizes a spatially homogeneous state of chemical species and generates patterns.
Since Turing's pioneering work, most studies on Turing patterns have been conducted within the framework of deterministic equations~\cite{kuramoto1984chemical,cross1993pattern}.

% Intro #2
Recently, pattern formation phenomena have also been observed in mesoscopic or microscopic systems
~\cite{huh2009pure, karig2018stochastic, schaffer2019stochastic, kohyama2020cell, fuseya2021nanoscale, yanagisawa2022cell, asaba2023growth, brandao2023learning, nishiguchi2025vortex}. 
In such systems, two key issues should be considered. 
The first issue is the effect of fluctuations, which are negligible in macroscopic systems. 
Within a broad context of stochastic pattern formation
~\cite{hohenberg1992effects, garcia1993effects, parrondo1996noise, jung1998noise, lindner2004effects, spagnolo2004noise, zarate2006hydrodynamic, sagues2007spatiotemporal, dobramysl2018stochastic}, 
the Turing mechanism has been studied
~\cite{butler2009robust,biancalani2010stochastic,cao2014stochastic,biancalani2017giant,rana2020precision}.
For example, recent studies have revealed that the Turing mechanism can amplify fluctuations of patterns even when the corresponding deterministic reaction-diffusion equations exhibit a spatially uniform stationary solution
~\cite{butler2009robust,biancalani2010stochastic,biancalani2017giant}. 
The second issue is the effect of spatial discreteness, 
namely that the spatial structure of the system has a discontinuous nature.
In particular, it has been reported that when the characteristic length of fluctuations is comparable to the microscopic cutoff length of a model,
continuum descriptions cannot be employed~\cite{kobayashi2023control,yanagisawa2025phase,sasa2025non-equilibrium}.
To the best of our knowledge, this second issue has not been examined in the context of the Turing mechanism.

% Intro #3
In this paper, we study the effect of spatial discreteness on the Turing mechanism in a stochastic reaction-diffusion model.
The model consists of $N$ lattice sites, each associated with a reaction vessel of volume $\Omega$.
Here, chemical reactions are described by Markov jump processes in each vessel, and diffusion is modeled as hopping between neighboring vessels.
As a representative model of chemical reactions, we consider the Brusselator~\cite{nicolis1977self,qian2002concentration,nguyen2018phase}. 
By performing a standard system-size expansion in $\Omega$~\cite{gardiner2004handbook,kampen1992stochastic}, 
this stochastic model reduces to a set of diffusion-coupled deterministic Brusselator equations in the limit $\Omega \to \infty$. 
This deterministic model is known to exhibit a Turing instability for certain parameter values. 
Thus, it is reasonable to expect that periodic patterns are observed 
in the stochastic model when the limit $\Omega \to \infty$ is considered, provided that the Turing instability condition is satisfied.
We show that this conjecture does not hold when the characteristic length is comparable to the lattice spacing.

% Intro #4
The main result of this paper is that the emergent pattern crucially depends on the order of limits among $\Omega \to \infty$, $t \to \infty$, and $N \to \infty$.
If we first take the limit $\Omega\to\infty$ for the stochastic model, the deterministic equations are obtained.
These equations exhibit spatially chaotic stationary solutions as $N \to \infty$, which are induced by spatial discreteness.
On the other hand, if we first take the limit $t\to\infty$ 
for the stochastic model with the same parameter values and a large fixed $\Omega$, 
the spectrum of patterns in the steady state scales with the system size $N$.
More precisely, we find that the peak value of the spectrum is proportional to $N$ in the limit $N \to \infty$ for fixed large $\Omega/N$.
This scaling indicates that, in this limit, a periodic pattern emerges in the stochastic system.
To summarize these results, taking the limits in the order $\Omega \to \infty$, $t \to \infty$, and $N \to \infty$ leads to spatial chaos, 
whereas taking them in the order $t \to \infty$, $\Omega \to \infty$, and $N \to \infty$ yields a periodic pattern.
This novel behavior arises from the spatial discreteness in the stochastic model.

\begin{figure}
    \centering
    \includegraphics[width=\linewidth]{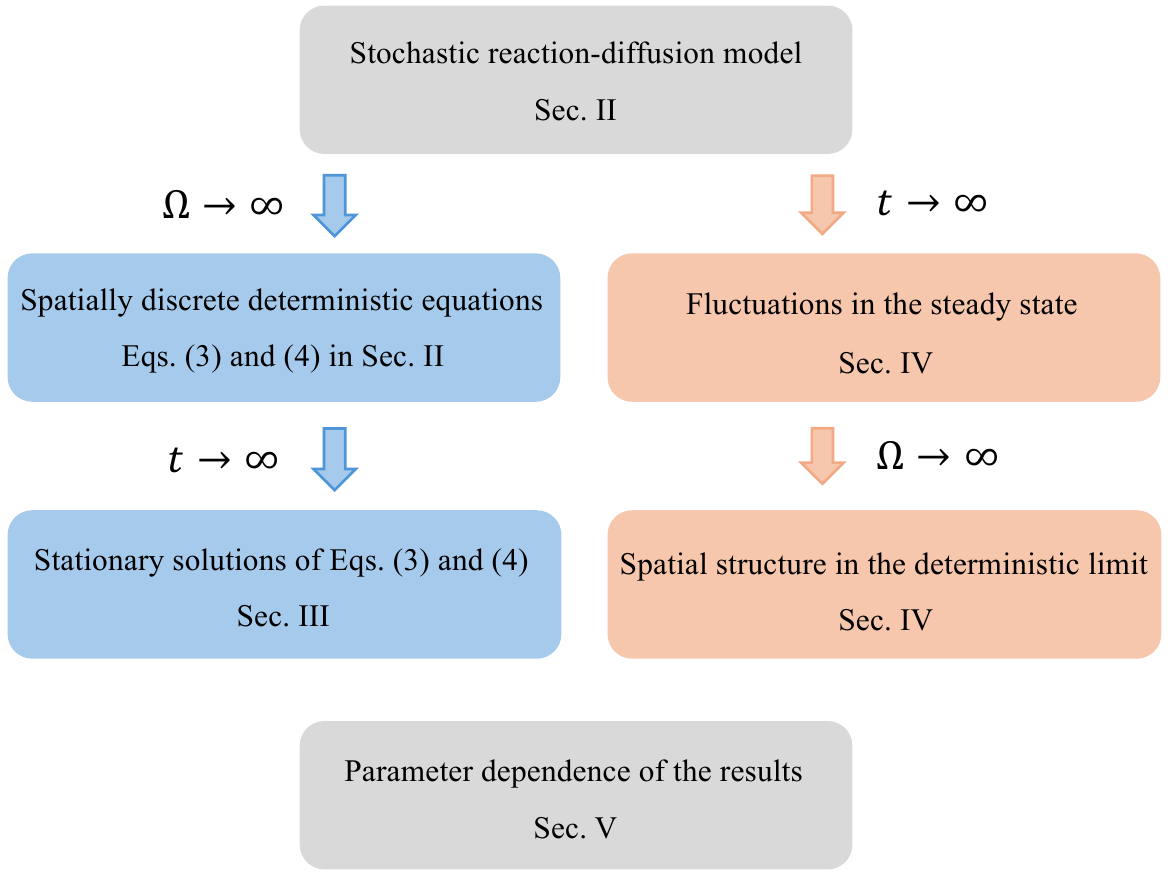}
    \caption{Graphical summary of this paper.}
    \label{summary}
\end{figure}

% Intro #5
This paper is organized as follows. 
In Sec.~\ref{setup}, we introduce a stochastic reaction-diffusion model and the corresponding spatially discrete deterministic equations. 
We then explain the condition under which the behavior of these discrete deterministic equations differs qualitatively from that of conventional reaction-diffusion equations.
In Sec.~\ref{DISC}, we numerically investigate the stationary solutions of the discrete deterministic equations. 
In Sec.~\ref{NIO}, we analyze the spatial structure of the stochastic model in the steady state 
and characterize it by means of a finite-size scaling analysis of the spectrum in the deterministic limit. 
In Sec.~\ref{others}, we examine how the results obtained in the previous sections depend on model parameters.
Section~\ref{concluding} is devoted to concluding remarks. 
Figure~\ref{summary} provides a graphical summary of this paper.

\section{Setup} \label{setup}

First, we consider a well-mixed chemical reaction system in a reaction vessel of volume $\Omega$.
We focus on a specific model known as the Brusselator~\cite{nicolis1977self,qian2002concentration,nguyen2018phase}, characterized by the chemical reactions
\begin{align}
    \begin{aligned}
        \ce{$A$ &<=>[k_1][k_2] $U$}, \\
        \ce{$U + B$ &->[k_3] $V$}, \\
        \ce{$2U + V$ &->[k_4] $3U$}.
    \end{aligned}
    \label{brusselator}
\end{align}
For any chemical species $Z \in \{A,B,U,V\}$, we use the same symbol $Z$ to denote the number of particles.
The concentration of $Z$ is represented by the small letter $z \; (=Z/\Omega)$.
Here, the concentrations $a$ and $b$ are fixed and we consider the dynamics of the concentrations $u$ and $v$.
For the remainder of the paper, we set $a=1$ and $k_{\mu}=1$ for $\mu=1, \dots ,4$ to make all quantities dimensionless, 
and we treat $b$ as a control parameter.

\begin{figure}
    \centering
    \includegraphics[width=0.9\linewidth]{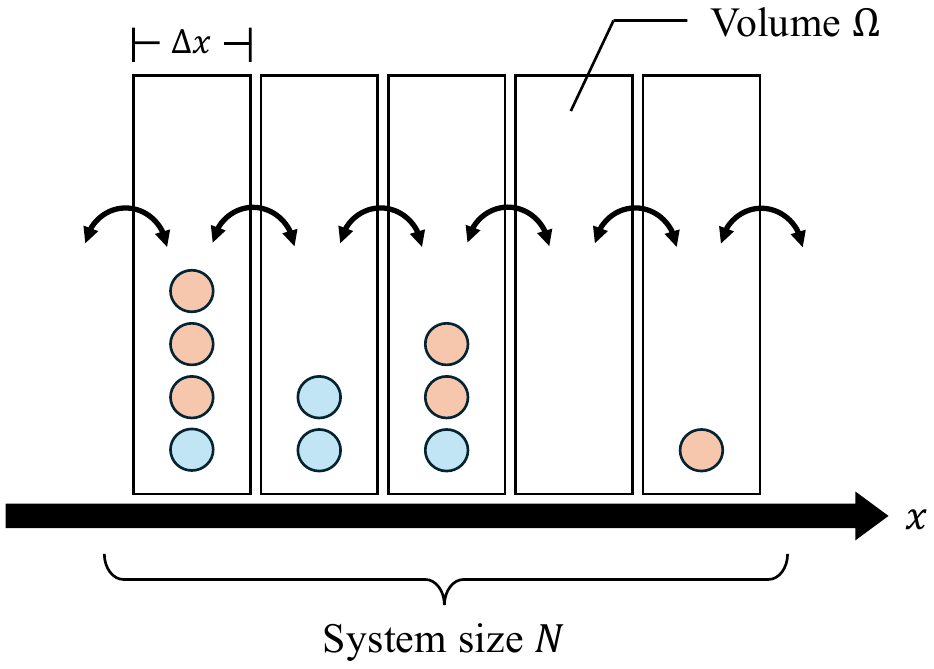}
    \caption{Schematic figure of a stochastic reaction-diffusion model.}
    \label{model}
\end{figure}

We now introduce a stochastic reaction-diffusion model~\cite{erban2020stochastic,yanagisawa2025phase}.
$N$ reaction vessels are arranged in one dimension along the $x$ direction (Fig.~\ref{model}).
Let $\Delta x$ denote the length of each vessel in the $x$ direction and $\mathcal{S}$ its cross-sectional area perpendicular to the $x$ direction.
The volume of each vessel is then given by $\Omega = \mathcal{S} \Delta x$.
Each vessel contains the chemical species $A,B,U,$ and $V$.
The numbers of particles $A$ and $B$ are fixed and take the same values in all vessels.
Let $U_i$ and $V_i$ denote the number of particles of species $U$ and $V$ in the $i$th vessel ($1\leq i\leq N$), respectively.
We assume that all vessels are well-mixed, and the state of the system is fully characterized by $(\vb*{U},\vb*{V})\equiv(U_1, \dots ,U_N,V_1, \dots ,V_N)$.
We also assume that chemical reactions occur only between particles in the same vessel and that particles can hop to adjacent vessels, with periodic boundary conditions imposed.

All chemical reactions and particle hoppings are described by the following Markov jump processes.
First, in the $i$th vessel, the state transitions corresponding to the chemical reactions in Eq.~\eqref{brusselator} are 
$U_i \to U_i+1, \; U_i \to U_i-1, \; (U_i,V_i) \to (U_i-1,V_i+1),$ and $(U_i,V_i) \to (U_i+1,V_i-1)$, with transition rates
$W_1,W_2,W_3,$ and $W_4$, respectively, where
\begin{align}
    \begin{aligned}
        W_1(U_i) &= \Omega, \\
        W_2(U_i) &= U_i, \\
        W_3(U_i,V_i) &= b U_i, \\
        W_4(U_i,V_i) &= U_i (U_i-1) V_i / \Omega^2.
    \end{aligned}
\end{align}
Second, state transitions associated with particle hoppings of species $U$ and $V$ from the $i$th vessel, 
$(U_i,U_{i+\sigma})\to(U_i-1,U_{i+\sigma}+1)$ and $(V_i,V_{i+\sigma})\to(V_i-1,V_{i+\sigma}+1)$ with $\sigma\in\{1,-1\}$, 
occur at rates $w_u U_i$ and $w_v V_i$, respectively.
The constants $w_u$ and $w_v$ represent the hopping rate per particle for chemical species $U$ and $V$, respectively.
These transitions model a diffusion process.

We now consider the limit $\Omega\to\infty$ of the stochastic reaction-diffusion model with $a$ and $b$ fixed. 
In this limit, the time evolution of $u_i \equiv U_i/\Omega$ and $v_i \equiv V_i/\Omega$ is described by the following spatially discrete deterministic equations~\cite{gardiner2004handbook,kampen1992stochastic}:
\begin{align}
    d_t u_i &= 1 - (1+b)u_i + u_i^2v_i \nonumber \\
    & \hspace{20mm} + w_u (u_{i+1} -2u_i +u_{i-1}), \label{ODE_u} \\
    d_t v_i &= b u_i - u_i^2v_i + w_v (v_{i+1} -2v_i +v_{i-1}). \label{ODE_v}
\end{align}
Here, we define diffusion coefficients $D_u \equiv w_u(\Delta x)^2$ and $D_v \equiv w_v(\Delta x)^2$.
Using the smallest diffusion coefficient $D_{\min} \equiv \min\{D_u, D_v\}$ and the maximum reaction rate $k_{\max} \equiv \max\{k_{\mu} \mid \mu=1, \dots ,4\}$,
we introduce the smallest characteristic length of patterns
\begin{align}
    \xi \equiv \sqrt{\frac{D_{\min}}{k_{\max}}},
\end{align}
which is referred to as the Kuramoto length~\cite{kampen1992stochastic,kuramoto1974effects}.
We then introduce the following important dimensionless quantity $\Xi$, which characterizes the extent of spatial continuity of the model:
\begin{align}
    \Xi \equiv \frac{\xi}{\Delta x}.
\end{align}
When the condition $\Xi\gg 1$ is satisfied, Eqs.~\eqref{ODE_u} and \eqref{ODE_v} can be approximated by the reaction-diffusion equations
\begin{align}
    \partial_t u &= 1 - (1+b)u + u^2v + D_u\partial_x^2 u, \label{PDE_u} \\
    \partial_t v &= b u - u^2v + D_v\partial_x^2 v, \label{PDE_v}
\end{align}
where we identify $u(x,t)\equiv u_i(t)$ and $v(x,t)\equiv v_i(t)$ with $x=i\Delta x$.
In contrast, in the case $\Xi=1$, solutions of Eqs.~\eqref{ODE_u} and \eqref{ODE_v} 
are found to be qualitatively different from those of Eqs.~\eqref{PDE_u} and \eqref{PDE_v}, as will be discussed later.

It is known that Eqs.~\eqref{PDE_u} and \eqref{PDE_v} exhibit a Turing instability~\cite{turing1952the,kuramoto1984chemical}. 
A linear stability analysis shows that this instability occurs when
\begin{align}
    b > \qty(1+\sqrt{\frac{D_u}{D_v}})^2.
\end{align}
A similar analysis of Eqs.~\eqref{ODE_u} and \eqref{ODE_v} shows that 
the spatially uniform solution becomes unstable to perturbations with certain discrete wavenumbers when
\begin{align}
    b > b_T \equiv \qty(1+\sqrt{\frac{w_u}{w_v}})^2.
    \label{turing_condition}
\end{align}
Throughout this work, we set $w_u/w_v=0.04$, for which $b_T=1.44$.

The aim of this paper is to characterize Turing patterns that arise in the stochastic reaction-diffusion model with $\Xi=1$.
In particular, we investigate how fluctuations affect Turing pattern formation by comparing the stationary solutions of the discrete deterministic equations 
with the spatial structure of the stochastic model in the steady state.
In the former case, we take the limits in the order $\Omega\to\infty$, $t\to\infty$, and $N\to\infty$, 
whereas in the latter case, we take them in the order $t\to\infty$, $\Omega\to\infty$, and $N\to\infty$.
We set $b=1.95$ so that the Turing instability condition Eq.~\eqref{turing_condition} is satisfied. 

\section{Stationary solutions of the discrete deterministic equations} \label{DISC}

In this section, we consider the case where the limit $\Omega\to\infty$ is taken first, followed by $t\to\infty$ and $N\to\infty$.
We numerically compute the stationary solutions of Eqs.~\eqref{ODE_u} and \eqref{ODE_v}. 
As the initial conditions, we take the spatially uniform solution $(u_0, v_0) = (1, b)$ and add random perturbations at each site by setting
$u_i = u_0 (1 + \xi_i)$ and $v_i = v_0 (1 + \eta_i)$, where $\xi_i$ and $\eta_i$ are uniformly sampled from the interval $[-0.01, 0.01]$ for $i = 1, \dots, N$.
The time evolution is computed using the Euler method. 
The time step is chosen as $\Delta t = 0.05 / w_v$.
We define a numerically obtained stationary solution as the configuration at the first time when $\max_i \{|\partial_t u_i|, |\partial_t v_i|\}$ falls below a prescribed threshold, which we set to $10^{-6}$.

\begin{figure}
    \centering
    \includegraphics[width=0.9\linewidth]{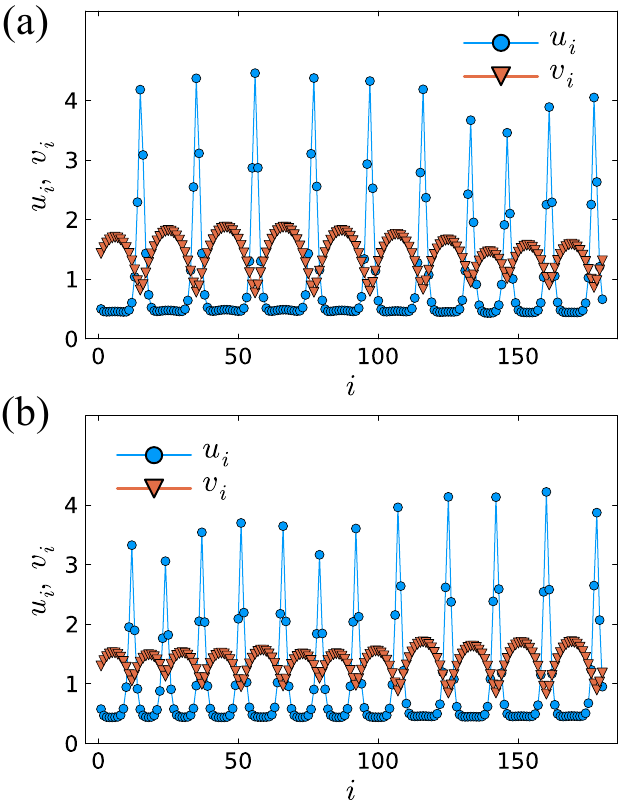}
    \caption{Examples of stationary solutions of Eqs.~\eqref{ODE_u} and \eqref{ODE_v} with $N = 180$. 
    The initial conditions in (a) and (b) are different.}
    \label{ODE_sol}
\end{figure}

Figure~\ref{ODE_sol} shows two examples of stationary solutions for $N = 180$, obtained from different initial conditions.
These stationary solutions exhibit irregular patterns that strongly depend on initial conditions.

\begin{figure}
    \centering
    \includegraphics[width=0.9\linewidth]{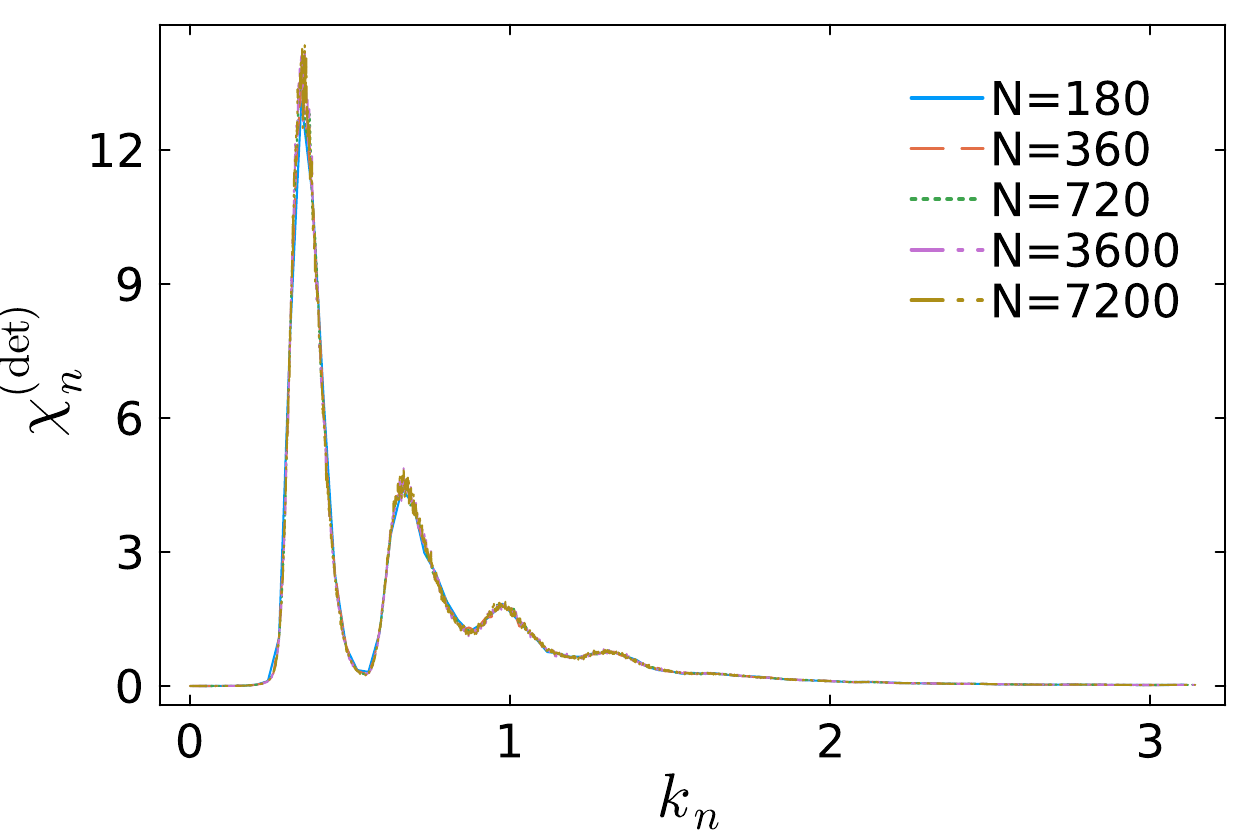}
    \caption{Spectrum $\chi_n^{\mathrm{(det)}}$ of the stationary patterns 
    for several system sizes $N$. 
    For each $N$, the spectrum is calculated by averaging $|\tilde{u}_n|^2$ over $1000$ stationary solutions 
    obtained from different initial conditions.}
    \label{ODE_spec_u_Wu1}
\end{figure}

To quantitatively characterize these stationary patterns, we introduce the spectrum $\chi_n^{\mathrm{(det)}}$ of the pattern $(u_i)_{i=1}^N$. 
Using the discrete Fourier transform
\begin{equation}
    \tilde{u}_n = \sum_{j=0}^{N-1} u_j e^{-i k_n j},
\end{equation}
where $k_n\equiv 2\pi n/N$,
the spectrum is defined as
\begin{equation}
    \chi_n^{\mathrm{(det)}} \equiv \frac{1}{N} \big\langle |\tilde{u}_n|^2 \big\rangle_{\mathrm{st}}.
\end{equation}
Here, $\langle \cdots \rangle_{\mathrm{st}}$ denotes an average over stationary solutions of Eqs.~\eqref{ODE_u} and \eqref{ODE_v}.
In practice, for each system size $N$, we prepare $1000$ initial conditions and 
compute the average of $|\tilde{u}_n|^2$ over the obtained stationary solutions.
The resulting spectrum is shown in Fig.~\ref{ODE_spec_u_Wu1}. 
We see that the peak value of the spectrum does not depend on $N$, and the peak width does not shrink with increasing $N$. 
This confirms that the stationary patterns indeed have no spatial periodicity or quasi-periodicity.

To further investigate the irregularity of the stationary patterns, we consider the stationary conditions of Eqs.~\eqref{ODE_u} and \eqref{ODE_v}, 
namely $\partial_t u_i = 0$ and $\partial_t v_i = 0$, and regard these equations as recurrence relations that determine the stationary patterns.
We then compute the maximum Lyapunov exponent of the resulting discrete dynamical system. 
Introducing
\begin{align}
    p_i &\equiv u_{i+1} - u_i, \\
    q_i &\equiv v_{i+1} - v_i,
\end{align}
we rewrite $\partial_t u_i=0$ and $\partial_t v_i=0$ as the four-dimensional map
\begin{align}
    \begin{aligned}
        u_{i+1} &= u_i + p_i, \\
        v_{i+1} &= v_i + q_i, \\
        p_{i+1} &= p_i - \frac{1}{w_u} \big\{1-(1+b)(u_i+p_i) \\
        &\hspace{34mm} +(u_i+p_i)^2(v_i+q_i)\big\}, \\
        q_{i+1} &= q_i - \frac{1}{w_v} \left\{b(u_i+p_i)-(u_i+p_i)^2(v_i+q_i)\right\},
    \end{aligned}
\label{reccurent_formula_nonlinear}
\end{align}
subject to periodic boundary conditions such as $u_{N+1} = u_1$. 

Here, we consider a periodic orbit $(u_i^*, v_i^*, p_i^*, q_i^*)_{i=1}^N$ that satisfies Eq.~\eqref{reccurent_formula_nonlinear}.
As such an orbit, we choose one of the stationary solutions obtained by direct numerical integration of Eqs.~\eqref{ODE_u} and \eqref{ODE_v}. 
Linearizing Eq.~\eqref{reccurent_formula_nonlinear} around this orbit by writing $u_i=u_i^*+\delta u_i, \ldots ,q_i=q_i^*+\delta q_i$,
we obtain the following linearized recurrence relations:
\begin{align}
    \begin{aligned}
    \delta u_{i+1} &= \delta u_i + \delta p_i, \\
    \delta v_{i+1} &= \delta v_i + \delta q_i, \\
    \delta p_{i+1} &= \delta p_i - \frac{1}{w_u} \big\{ \qty(-(1+b)+2(u_i^*+p_i^*)(v_i^*+q_i^*)) \\
    &\quad \times(\delta u_i+\delta p_i) +(u_i^*+p_i^*)^2(\delta v_i+\delta q_i) \big\}, \\
    \delta q_{i+1} &= \delta q_i - \frac{1}{w_v} \big\{ \qty(b-2(u_i^*+p_i^*)(v_i^*+q_i^*)) \\
    &\quad \times(\delta u_i+\delta p_i) -(u_i^*+p_i^*)^2(\delta v_i+\delta q_i) \big\}.
    \end{aligned}
    \label{reccurent_formula}
\end{align}

We prepare two arbitrary unit vectors 
$\delta \vb*{x}'_1\equiv (\delta u'_1, \ldots , \delta q'_1)$ and $\delta \vb*{x}''_1\equiv (\delta u''_1, \ldots , \delta q''_1)$ 
with $|\delta\vb*{x}'_1|=|\delta\vb*{x}''_1|=1$, and apply the linear map Eq.~\eqref{reccurent_formula} to obtain $\delta \vb*{x}'_2$ and $\delta \vb*{x}''_2$. 
We then renormalize these vectors so that $|\delta\vb*{x}'_2|=|\delta\vb*{x}''_2|=1$, and by iterating this procedure over one period, 
we obtain vectors $\delta \vb*{x}_{N+1}'$ and $\delta \vb*{x}_{N+1}''$ again with unit norm. 
We confirm that $\delta \vb*{x}_{N+1}'$ and $\delta \vb*{x}_{N+1}''$ coincide, 
indicating that this procedure yields the Lyapunov (Floquet) vector. 

We then take this Lyapunov vector as the initial condition $\delta \vb*{x}_1$ with $|\delta \vb*{x}_1|=1$ for the following calculation.
Starting from $\delta \vb*{x}_1$, we compute $\delta \vb*{x}_2$ using Eq.~\eqref{reccurent_formula} and define the local expansion rate
\begin{align}
    \mu_1 \equiv |\delta \vb*{x}_2|.
\end{align}
After renormalizing $\delta \vb*{x}_2$ to a unit vector, we compute $\delta \vb*{x}_3$ and obtain $\mu_2$ in the same manner. 
Repeating this procedure over one period yields $\mu_i$ for $i = 1,\dots,N$. 
The maximum Lyapunov exponent of the dynamical system Eq.~\eqref{reccurent_formula_nonlinear} is then given by
\begin{align}
    \lambda_{\max}^N = \frac{1}{N} \sum_{i=1}^{N} \log \mu_i. \label{max_Lyapunov_formula}
\end{align}

\begin{figure}
    \centering
    \includegraphics[width=0.9\linewidth]{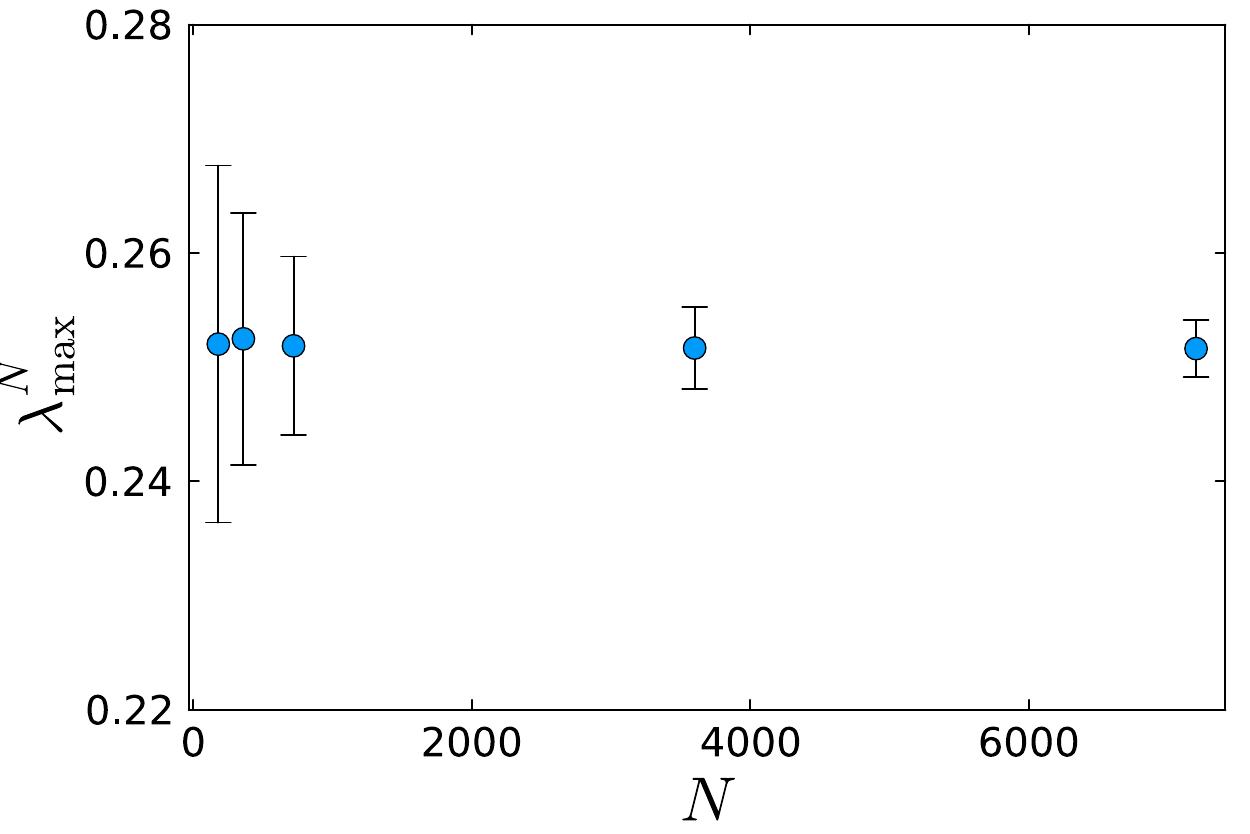}
    \caption{$N$-dependence of the distribution of maximum Lyapunov exponent.
    For each system size $N$, the average over $1000$ orbits is plotted.
    The error bars represent the standard deviation.}
    \label{max_Lyapunov}
\end{figure}

Because the value of the maximum Lyapunov exponent $\lambda_{\max}^N$ depends on the orbit $(u_i^*, v_i^*, p_i^*, q_i^*)_{i=1}^N$,
we compute $\lambda_{\max}^N$ for $1000$ stationary solutions, which are the same as those used to calculate the spectrum in Fig.~\ref{ODE_spec_u_Wu1}.
Figure~\ref{max_Lyapunov} shows their average for several system sizes $N$, with error bars representing the standard deviation.
The result shows that the average value is positive and that the standard deviation decreases with increasing $N$.
This suggests that the maximum Lyapunov exponent takes a positive value for almost all orbits as $N\to\infty$.
Therefore, we conclude that the stationary solutions of Eqs.~\eqref{ODE_u} and \eqref{ODE_v} exhibit spatial chaos in the limit $N\to\infty$. 

Note that spatially chaotic solutions of partial differential equations have been reported in previous studies \cite{coullet1987nature,bazhenov1996time}.
In contrast, the spatial chaos observed in the present study arises due to the discreteness of the space.
To the best of our knowledge, such discreteness-induced spatial chaos has not been reported previously.

\section{Deterministic limit of the steady state of the stochastic model} \label{NIO}

In this section, we consider another order of limits: we first take the limit $t \to \infty$, followed by $\Omega \to \infty$ and $N\to\infty$. 
We return to the stochastic reaction-diffusion model and analyze its spatial structure in the steady state. 
All parameter values are chosen to be identical to those in Sec. \ref{DISC}. 
The stochastic model is simulated using the Gillespie algorithm \cite{gillespie2007stochastic}.

\begin{figure}
    \centering
    \includegraphics[width=0.9\linewidth]{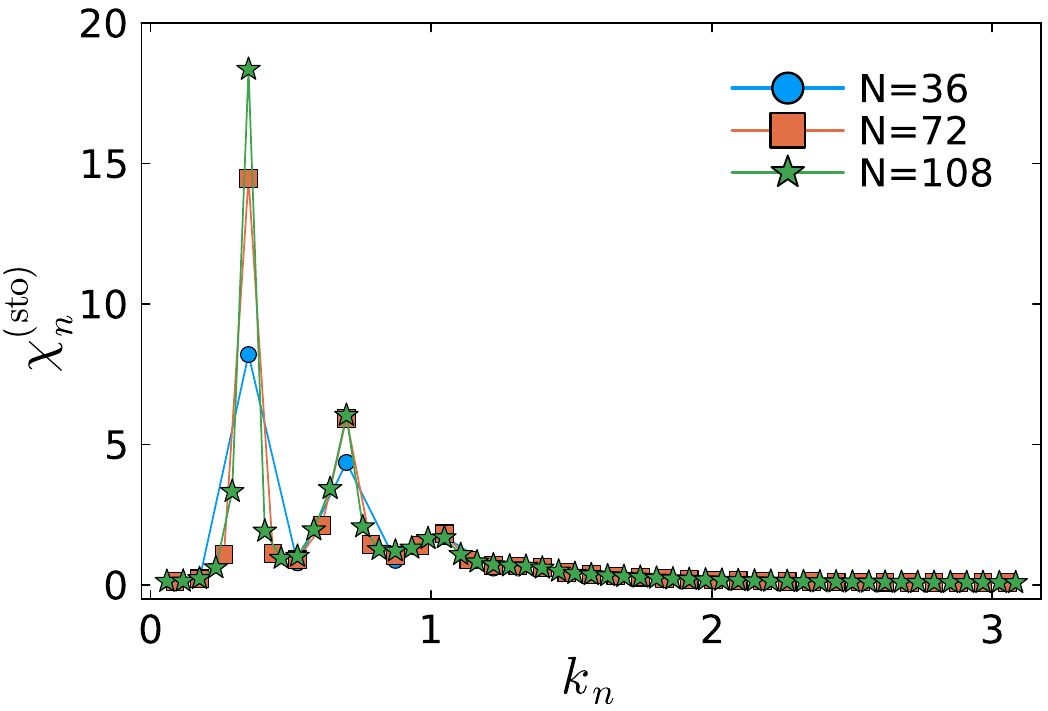}
    \caption{Spectrum $\chi_n^{(\mathrm{sto})}$ in the steady state of the stochastic model 
    for $\Omega=40$ with $N = 36, 72,$ and $108$.}
    \label{spec_b1_95_N}
\end{figure}

\begin{figure}
    \centering
    \includegraphics[width=0.9\linewidth]{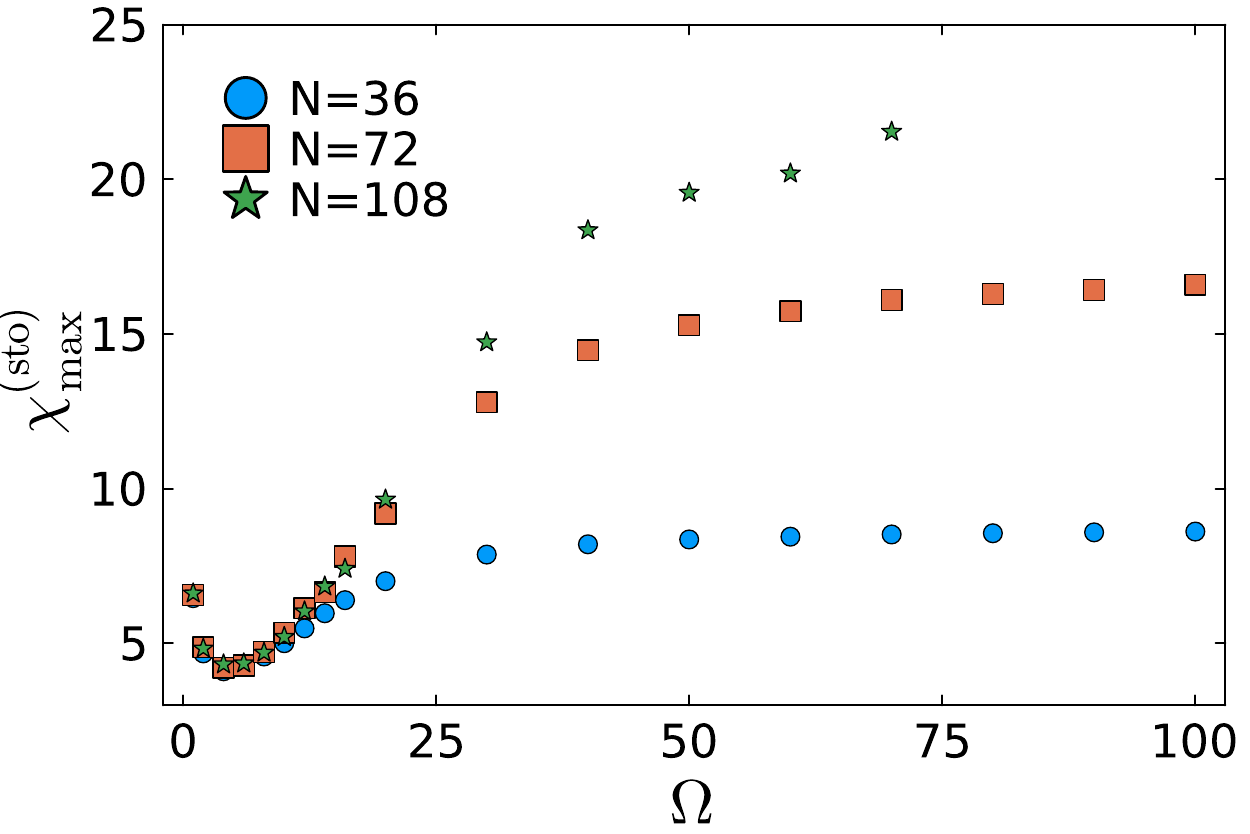}
    \caption{$\Omega$-dependence of the spectrum peak value $\chi_{\max}^{\mathrm{(sto)}}$ 
    for $N=36, 72,$ and $108$.}
    \label{spec_b1_95}
\end{figure}

To quantitatively characterize the patterns formed in the stochastic model, 
we introduce the spectrum
\begin{align}
    \chi_n^{(\mathrm{sto})} \equiv \frac{1}{N} \big\langle |\tilde{u}_n|^2 \big\rangle, \label{def_spectrum}
\end{align}
where $\langle \cdots \rangle$ denotes the statistical average in the steady state.
Figure~\ref{spec_b1_95_N} shows the spectrum for $\Omega=40$ with $N = 36, 72,$ and $108$.
Here, the spectrum is estimated by taking the time average over the configurations from $t = 16383$ to $t = 32767$.
We observe that the peak value of the spectrum, $\chi_{\max}^{\mathrm{(sto)}}$, increases with $N$.
Figure~\ref{spec_b1_95} shows how this peak value depends on $\Omega$ for each $N$. 
As $\Omega$ increases, the peak value saturates to a certain value for each fixed $N$, 
and this saturation value increases with $N$.

\begin{figure}
    \centering
    \includegraphics[width=0.9\linewidth]{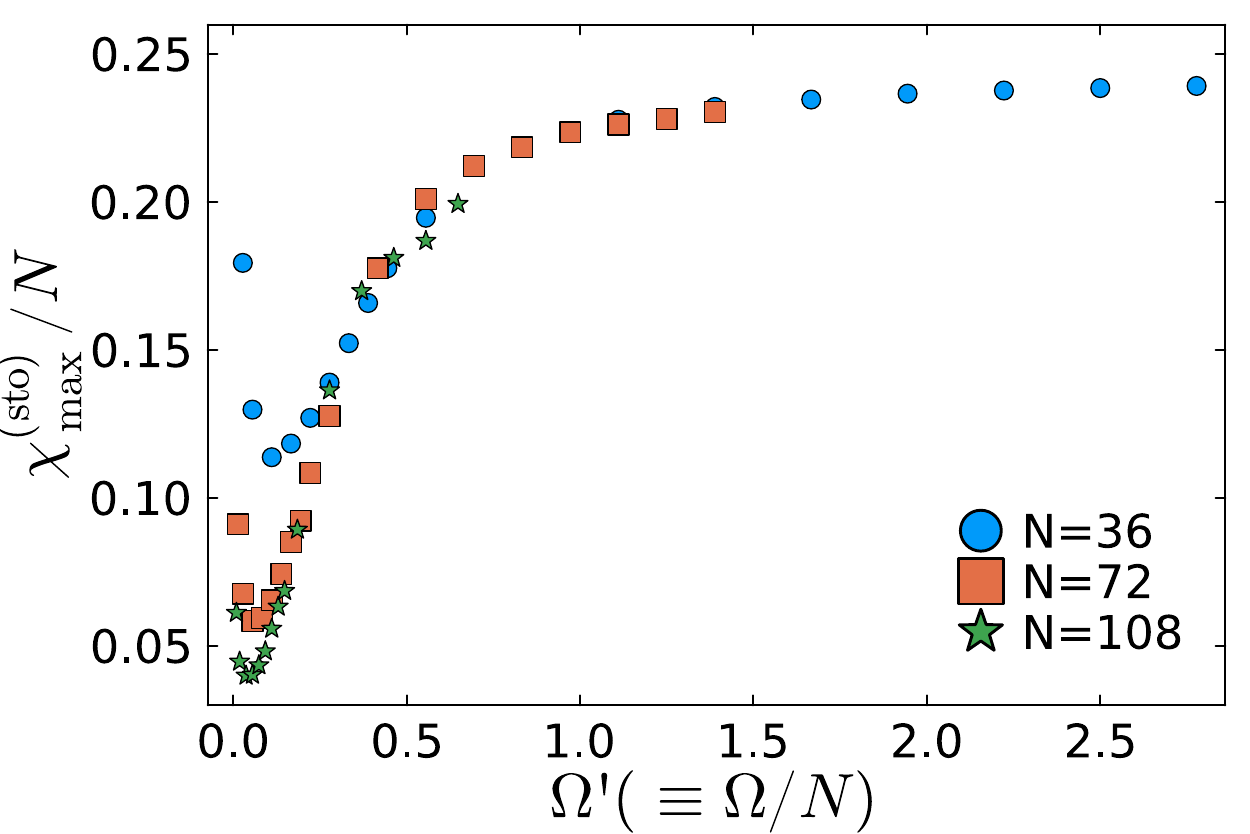}
    \caption{Finite-size scaling of the spectrum peak data shown in Fig.~\ref{spec_b1_95}.}
    \label{spec_b1_95_scaling}
\end{figure}

We next examine the spatial structure of the stochastic model in the deterministic limit $\Omega \to \infty$. 
To this end, we perform a finite-size scaling analysis of the spectrum peak value with respect to $N$. 
Figure~\ref{spec_b1_95_scaling} shows the data from Fig.~\ref{spec_b1_95} replotted with the horizontal axis rescaled as $\Omega' \equiv \Omega/N$
and the vertical axis as $\chi_{\max}^{\mathrm{(sto)}} / N$.
We observe that, for sufficiently large $\Omega'$, the data for different system sizes collapse onto a single curve. 
Since, for fixed $N$, the limit $\Omega \to \infty$ corresponds to $\Omega' \to \infty$, 
the observed data collapse implies that the quantity $\chi_{\max}^{\mathrm{(sto)}}/N$ approaches a finite value as $\Omega \to \infty$. 
Subsequently taking the limit $N \to \infty$, we thus conclude that $\chi_{\max}^{\mathrm{(sto)}}$ diverges linearly with $N$.
Therefore, the stationary state of the stochastic model becomes spatially periodic in the deterministic limit.

In the present analysis, we first take the limit $t \to \infty$ for finite $\Omega$ and then take $\Omega \to \infty$.
The results in this section thus demonstrate that the pattern becomes periodically ordered due to small fluctuations, giving rise to fluctuation-induced spatial order.
Although it is known that small noise can turn a chaotic time series into a periodic orbit \cite{matsumoto1983noise},
the relationship between this phenomenon and our case remains to be elucidated.

\section{Results for other parameter values} \label{others}

\begin{figure}
    \centering
    \includegraphics[width=0.9\linewidth]{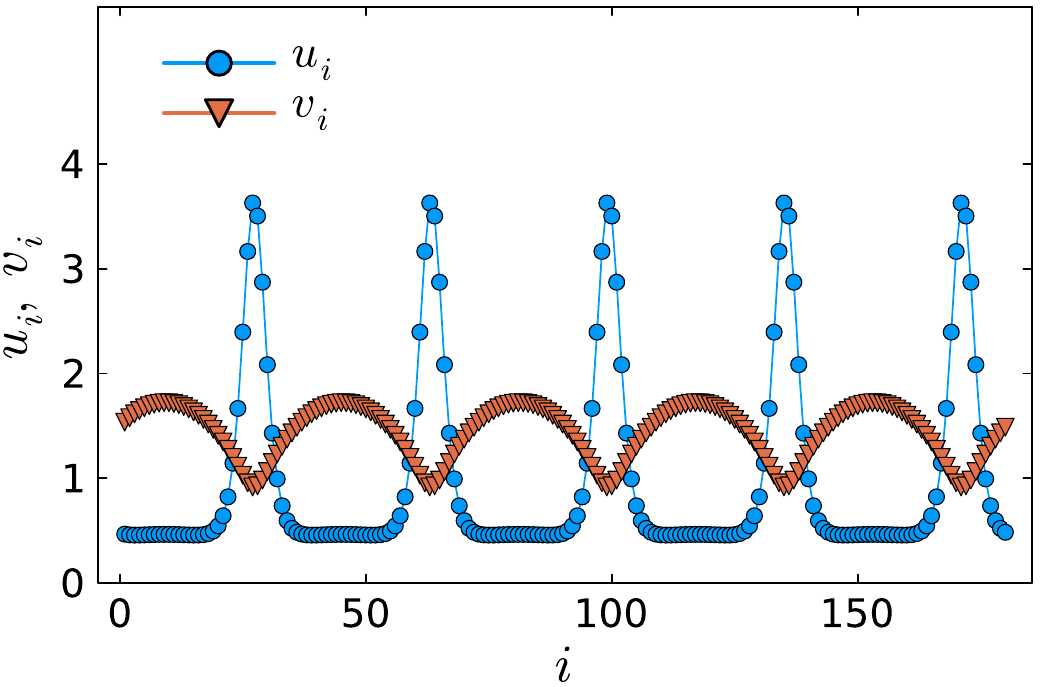}
    \caption{Example of a stationary solution of Eqs.~\eqref{ODE_u} and \eqref{ODE_v} for $\Xi=2$.
    All other parameter values are the same as in Fig.~\ref{ODE_sol}.}
    \label{ODE_sol_Xi2}
\end{figure}

In this section, we investigate how the results obtained in the previous sections depend on the model parameters.
We first examine the case $\Xi = 2$, 
for which the system is closer to a partial differential equation description.
Figure~\ref{ODE_sol_Xi2} shows an example of a stationary solution of Eqs.~\eqref{ODE_u} and \eqref{ODE_v} for this case, with $N=180$.
In contrast to the case $\Xi=1$ shown in Fig.~\ref{ODE_sol}, we find that the stationary solution is periodic in the present case.
This result supports the conclusion that the spatially chaotic stationary solutions obtained for $\Xi = 1$ are induced by spatial discreteness.
We note that periodic stationary solutions are not unique and that their wavelength depends on random initial conditions.

\begin{figure}
    \centering
    \includegraphics[width=0.9\linewidth]{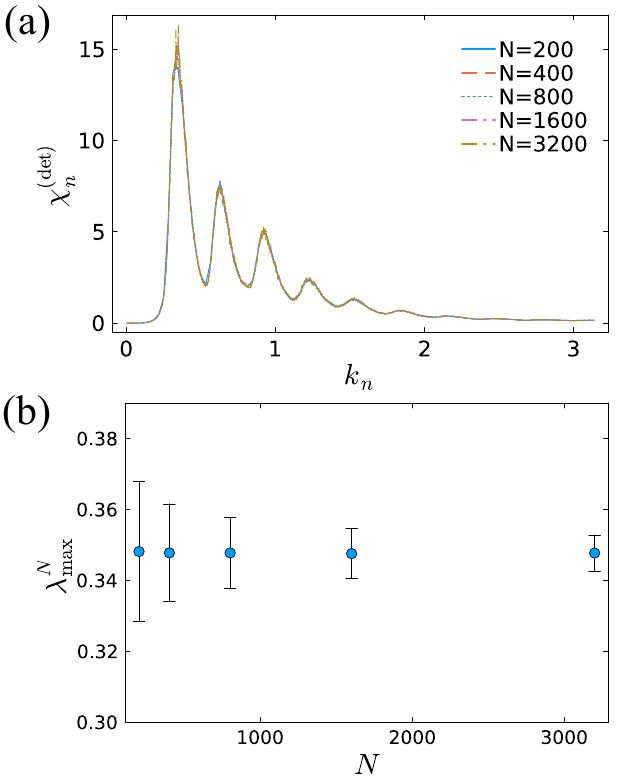}
    \caption{Analyses of the stationary solutions of the discrete deterministic equations~\eqref{ODE_u} and \eqref{ODE_v} for $b=2.5$.
        (a) Spectrum $\chi_n^{\mathrm{(det)}}$ of the stationary solutions for several system sizes $N$.
        (b) $N$-dependence of the distribution of maximum Lyapunov exponent.
        The error bars represent the standard deviation.}
    \label{ODE_b2_5}
\end{figure}

\begin{figure}
    \centering
    \includegraphics[width=0.9\linewidth]{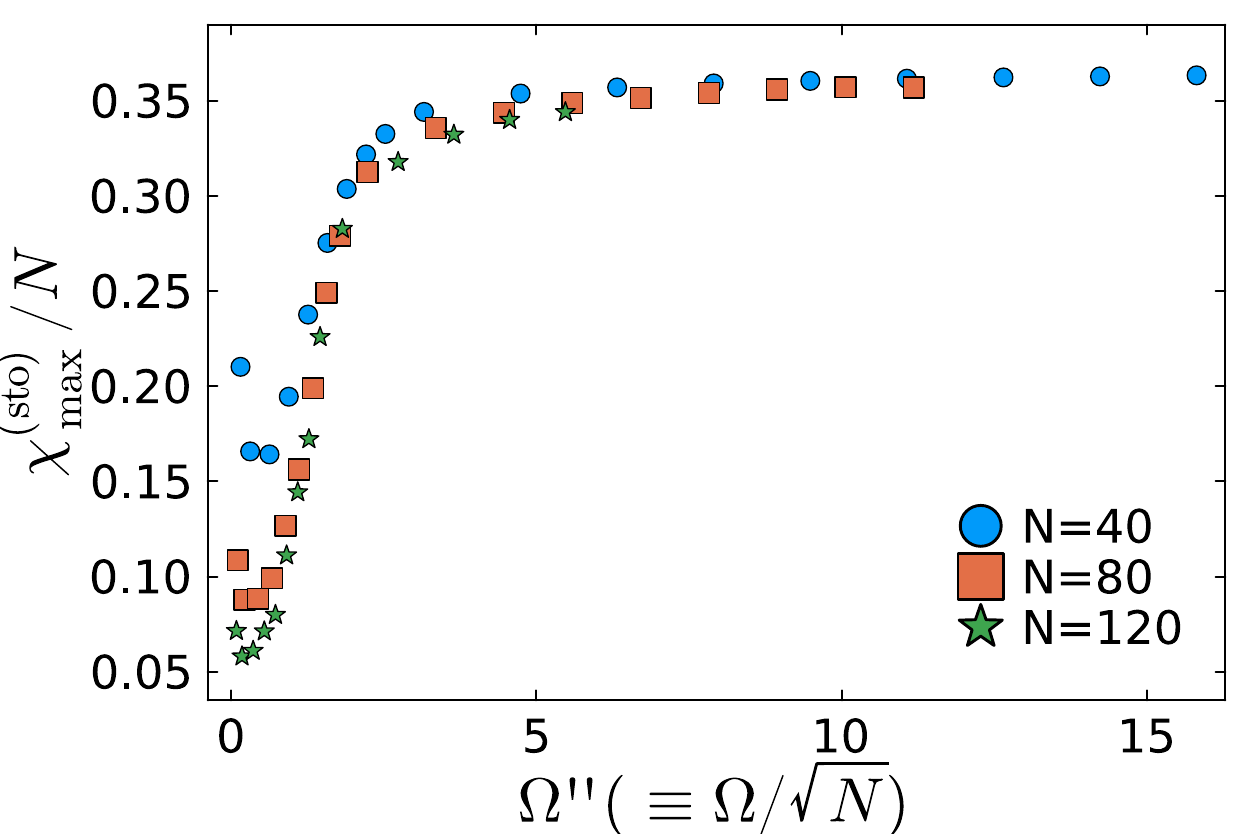}
    \caption{Finite-size scaling of the spectrum peak value $\chi_{\max}^{\mathrm{(sto)}}$ in the steady state for $b=2.5$.}
    \label{spec_b2_5_scaling}
\end{figure}

Next, to study the $b$-dependence, we focus on $b = 2.5$ and repeat the analyses of Secs.~\ref{DISC} and \ref{NIO}, now setting $\Xi = 1$.
When we first take the limit $\Omega \to \infty$ and then $t \to \infty$, 
the stationary solutions of Eqs.~\eqref{ODE_u} and \eqref{ODE_v} are again found to be spatially chaotic, 
as in the case $b = 1.95$ (see Fig.~\ref{ODE_b2_5}). 
On the other hand, when we first take the limit $t \to \infty$ and then $\Omega \to \infty$, 
we find that a different scaling exponent is obtained in the finite-size scaling of the spectrum peak value. 
More precisely, for sufficiently large $\Omega'' \equiv \Omega / \sqrt{N}$,
the data for different $N$ collapse onto a single curve when $\chi_{\max}^{\mathrm{(sto)}} / N$ is plotted as a function of $\Omega''$ (see Fig.~\ref{spec_b2_5_scaling}).

Even in this case, fixing $N$ and taking the limit $\Omega \to \infty$ implies $\Omega'' \to \infty$, and the data collapse for large $\Omega''$ 
shows that $\chi_{\max}^{\mathrm{(sto)}}$ diverges linearly with $N$ in this limit.
Thus, the pattern in the steady state is again spatially periodic in the deterministic limit. 
Therefore, we also observe fluctuation-induced spatial order for $b = 2.5$, 
i.e., spatially chaotic patterns become periodically ordered due to small fluctuations.
A more detailed understanding of the scaling exponents and their universality is left for future work.

\section{Concluding remarks} \label{concluding}

In this paper, we analyzed the Turing patterns that arise in a system where spatial discreteness plays a significant role. 
By comparing the two limiting procedures 
--- one with the limits taken in the order $\Omega \to \infty$, $t \to \infty$, $N \to \infty$, and another with $t \to \infty$, $\Omega \to \infty$, $N \to \infty$ ---
we examined how fluctuations affect Turing pattern formation.
We found that, in the former case, the stationary solutions of the discrete deterministic equations~\eqref{ODE_u} and \eqref{ODE_v} are spatially chaotic, 
whereas, in the latter case, the patterns become spatially periodic. 
The former behavior represents discreteness-induced spatial chaos, in which irregular patterns arise due to spatial discreteness,
while the latter behavior represents fluctuation-induced spatial order, 
in which the patterns become periodically ordered due to small fluctuations.

Turing patterns in systems with spatial discreteness, as studied here,
are expected to be important for understanding experimental results on reaction-diffusion systems at mesoscopic or microscopic length scales~\cite{karig2018stochastic, kohyama2020cell, fuseya2021nanoscale, asaba2023growth}. 
In such reaction-diffusion systems, the characteristic length of patterns can become comparable to the length scale of the system's constituents.
Using such systems, one may experimentally observe the spatially chaotic patterns induced by spatial discreteness, as well as the fluctuation-induced spatial ordering of these patterns found in the present work.
Such observations would shed light on the limitations of macroscopic pattern formation theories 
and highlight the need for new theories of microscopic pattern formation phenomena, 
with the present work serving as a guiding example.

% acknowledgment
The authors thank Masato Itami, Ryudo Suzuki, and Jann van der Meer for their helpful discussion.
This study was supported by JSPS KAKENHI Grant Numbers JP22K13975, JP23K22415, JP25K00923, JP25H01975,
and JST SPRING Grant Number JPMJSP2110.

\end{document}